\documentclass[mathleft]{an}

\usepackage{graphicx}
\usepackage{times}
\usepackage{natbib}
\bibpunct{(}{)}{;}{a}{}{,}
\newcommand{\sm}{M$_{\odot}$}
\newcommand{\sr}{R$_{\odot}$}

\newcommand{\sz}{Z$_{\odot}$}
\newcommand{\msol}{M$_{\odot}$}
\newcommand{\ergs}{$~{\rm erg~s^{\scriptscriptstyle -1}}$}

\newcommand{\kms}{\ensuremath{{\rm km~s}^{-1}}}

\newcommand{\snxray}{SN~2008D/XRT~080109}
\newcommand{\snaj}{SN~2006aj/XRF~060218}
\newcommand{\snbw}{SN~1998bw/GRB~980425}

\newcommand{\OxySeven}{\ion{O}{1}~$\lambda$7774}
\newcommand{\MgOne}{[\ion{Mg}{1}~$\lambda4571$}

\newcommand{\synNi}{$^{56}$\rm{Ni}}

\begin{document}

\Yearpublication{2011}%
\Yearsubmission{2011}%
\Month{06}%
\Volume{332}%
\Issue{5}%

\title{Stellar Forensics with the Supernova-GRB Connection \\ (Ludwig-Biermann Award Lecture 2010) }

\author{Maryam Modjaz\inst{1}\fnmsep\thanks{ \email{mmodjaz@gmail.com}\newline}
}
\titlerunning{Stellar Forensics with the SN-GRB Connection}
\authorrunning{Maryam Modjaz}
\institute{Hubble Fellow, Columbia University, Pupin Physics Laboratories, MC 5247,
550 West 120th Street,
New York, NY 10027, USA 
}

\received{2011, Apr 1}
\accepted{2011, May 2}


\abstract{%
Long-duration gamma-ray bursts (GRBs) and Type Ib/c Supernovae (SNe
Ib/c) are amongst nature's most magnificent explosions. While GRBs launch relativistic jets, SNe Ib/c are core-collapse explosions whose progenitors have been stripped of their hydrogen and
helium envelopes. Yet for over a decade, one of the key outstanding
questions is which conditions lead to each kind of explosion and death in massive stars. Determining the fates of massive stars is not only a vibrant topic in itself, but also impacts using GRBs as star formation
indicators over distances of up to 13 billion light-years and for mapping
the chemical enrichment history of the universe. This article reviews a number of comprehensive observational studies that probe the progenitor environments, their metallicities and the
 explosion geometries of SN with and without GRBs, as well as the emerging field of SN environmental studies.
   Furthermore, it discusses \snxray\, which was discovered serendipitously with the {\it Swift} satellite via its X-ray emission from shock breakout, and which has
 generated great interest amongst both observers and theorists while
 illustrating a novel technique for stellar forensics. The article concludes
 with an outlook on how the most promising venues of research - with the many existing and upcoming large-scale surveys such as the Palomar Transient Factory and  LSST -
 will shed new light on the diverse deaths of massive stars.}
\maketitle

\section{Introduction: The Importance of Stellar Forensics}\label{intro_sec} 

\begin{figure*}[!ht]
\includegraphics[scale=0.55,angle=-90]{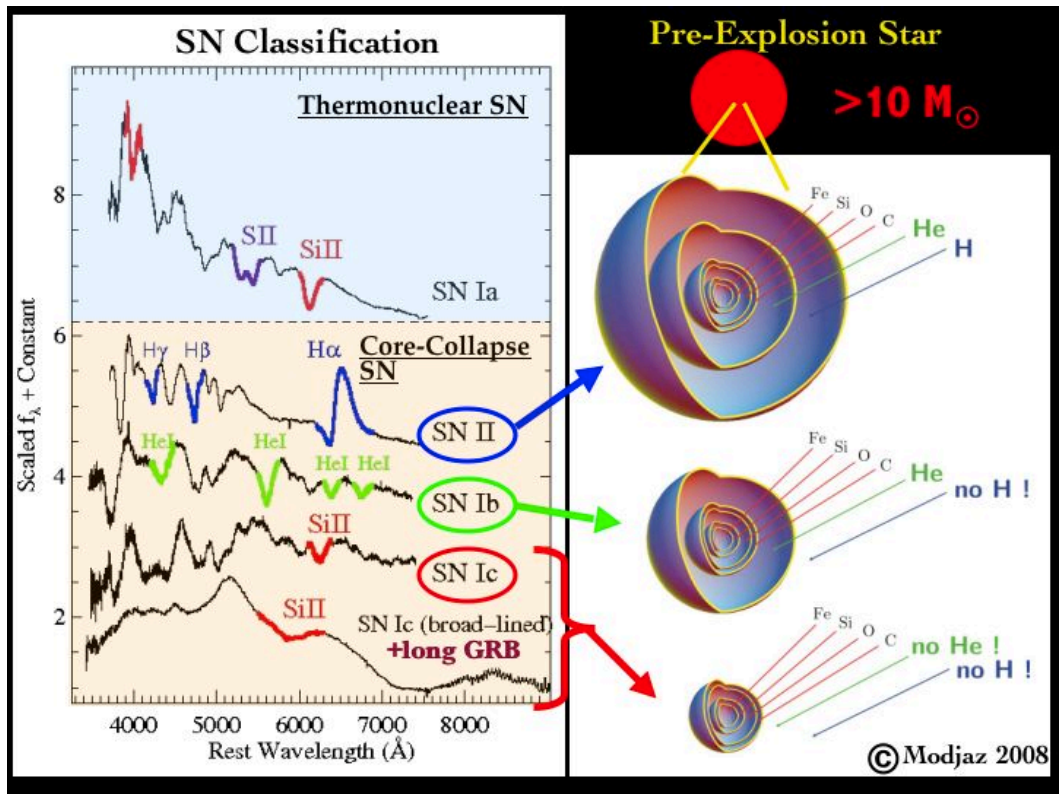}

\caption{Mapping between different types of core-collapse SNe ({\it left}) and their corresponding progenitor stars ({\it right}). {\it Left}: Representative observed spectra of different types of SNe. Broad-lined SN Ic are the only type of SNe seen in conjunction with GRBs. Not shown are some of the other H-rich SN members (SNe IIn and very luminous SNe). {\it Right}: Schematic drawing of massive ($\ge$ 8$-10$ \sm) stars before explosion, with different amounts of intact outer layers, showing the "onion-structure" of different layers of elements that result from successive stages of nuclear fusion during the massive stars' lifetimes (except for H). The envelope sizes are not drawn to scale; in particular, the outermost Hydrogen envelope at the top can be up to 100$-$1000 times larger than shown. Furthermore, many real massive stars rotate rapidly and are therefore oblate, as well as show much less chemical stratification due to convection and overshoot mixing (e.g., see review by \citealt{woosley02}) than is drawn here. The bottom star constitute the most stripped (or "naked") star, with a typical size of $\sim$~\sr , whose demise produces a SN Ic and sometimes, a SN Ic-bl and even more rarely, a SN Ic-bl accompanied by GRBs. One of the outstanding questions in the field is the dominant mechanism with which the outer H and He layers got removed. This figure can be downloaded at http://www.astro.columbia.edu/~mmodjaz/research.html  }
 \label{SNclass_fig}
\end{figure*}

Stripped supernovae (SNe) and long-duration Gamma-Ray Bursts (GRBs) are nature's most powerful explosions from massive stars. They energize and enrich the
ISM, and, like beacons, they are visible over large cosmological distances.
However, the exact mass and metallicity range of their
progenitors is not known, nor the detailed physics of the explosion (see reviews by \citealt{woosley06_rev}, and by \citealt{smartt09_rev}). Stripped-envelope SNe (i.e, SN IIb, Ib, Ic and Ic-bl, e.g., \citealt{uomoto85,wheeler90, clocchiatti96,filippenko97_review}) are core-collapse events whose massive progenitors have been stripped of progressively larger
amounts of their outermost H and He envelopes (Fig.~\ref{SNclass_fig}). In particular, broad-lined SNe~Ic (SNe~Ic-bl) are SNe Ic whose line widths approach 30,000 \kms\ around maximum light and whose optical spectra show no trace of H and He.

The exciting connection between long GRBs and
SNe~Ic-bl (for a review, see \citealt{woosley06_rev,hjorth11} and below) and the existence of SNe~Ic-bl without
observed GRBs, as well as that of GRBs that surprisingly lack SN signatures, raises the
question of what distinguishes a GRB progenitor from that of an
ordinary SN~Ic-bl with and without a GRB. 

Understanding the progenitors of SNe Ib/c and of GRB is important on a number of levels:

(1) \textbf{Stellar and High-Energy Astrophysics}: These stellar explosions leave behind extreme remnants, such as Black Holes, Neutron Stars, and
Magnetars, which in themselves are a rich set of phenomena studied over the full wavelength spectrum from Gamma-rays to radio. Ideally we would like to construct a map that connects the mass and make-up of a massive star to the kind of 
death it undergoes and to the kind of remnant it leaves behind. Furthermore, these stellar explosions are sources of gravitational waves and of neutrino emission, and specifically GRBs are leading candidate sites for high-energy cosmic ray acceleration (e.g., \citealt{waxman04}). Thus, it is of broad astrophysical importance to understand the specific progenitor and production conditions for different kinds of cosmic explosions. 

(2) \textbf{Chemical Enrichment History of Universe}: 
The universe's first- and
second-generation stars were massive. 
Since GRBs and SNe probably
contribute differently to the enrichment of heavy elements (e.g., \citealt{pruet06,nomoto06}), determining the fate of massive stars is
fundamental to tracing the chemical history of the universe.

(3) \textbf{Cosmology}: GRBs are beacons and can illuminate the early universe. Indeed, until recently, the object with the highest spectroscopic redshift was a GRB, 
GRB090423 at $z \sim$8.2  \citep{tanvir09,salvaterra09}, which means that this explosion occurred merely 630 million years after the Big Bang. 
Thus, a clear understanding of the stellar progenitors of SNe and GRBs is an essential foundation for using them as indicators of
star formation over cosmological distances.

Various progenitor channels have been proposed for stripped SNe and GRBs: either
single massive Wolf-Rayet (WR) stars with main-sequence (MS) masses of
$\ga$ 30 \sm\ that have experienced mass loss during the MS and WR
stages (e.g., \citealt{woosley93}), or binaries from lower-mass
He stars that have been stripped of their outer envelopes through
interaction (\citealt{podsiadlowski04,fryer07}, and references therein), possibly given rise to run-away stars as GRB progenitors (e.g., \citealt{cantiello07,eldridge11}). For long GRBs, the main models for a central engine that is powering the GRB include the popular collapsar model \citep{woosley93_grb,macfadyen99} and the magnetar model (e.g., \citealt{usov92}, for a good summary see \citealt{metzger10}), while rapid rotation of the pre-explosion stellar core appears to be a necessary ingredient for both scenarios.

Attempts to directly identify SN~Ib/c progenitors
in pre-explosion images obtained with the {\it Hubble Space Telescope} 
or ground-based telescopes have not yet been successful
(e.g., \citealt{gal-yam05,maund05}; \citealt{smartt09_rev}), and do not conclusively distinguish
 between the two suggested progenitor scenarios. However, the
progenitor non-detections of 10 SNe Ib/c strongly indicate that the
single massive WR progenitor channel (as we observe in the Local
Group) cannot be the only progenitor channel for stripped SNe \citep{smartt09_rev}. Similar pre-explosion imaging technique is not possible for GRB progenitors given the large distances at which they are observed.

Thus, in order to fully exploit the potential and power of SNe and GRBs, we have to first figure out their stellar progenitors and the explosions conditions that lead to the various kinds of stellar death in a massive star, in form of a "stellar forensics" investigation.
In the following review, we will be looking at a number of inferred physical properties of the progenitor and of the explosion in the hopes of finding those that set apart SN-GRB (Section~\ref{sngrb_sec}) from SNe without GRBs. They are geometry of the explosion (Section~\ref{asphericity_sec}), progenitor mass (Section~\ref{mass_sec}) and metallicity (Section~\ref{metallicity_sec}), while the role of binaries are discussed through-out, but not that of magnetic fields. In addition, I will discuss the exciting and emerging field of SN metallicity studies as a promising new tool to probe the progenitors of different kinds of SNe and transients, as well as the story of \snxray\ (Section~\ref{08d_sec}) that generated great interest amongst both observers and theorists while illustrating a novel technique for stellar forensics.

Necessarily, this review will not be complete given the page limit, and is driven by the interest and work of the author, so omissions and simplifications will necessarily arise. Furthermore, given the excellent reviews by e.g.,  \citet{woosley06_rev} and most recently, \citet{hjorth11}, I will concentrate on developments in the field since 2006 and in complimentary areas.  

\section{Solid Cases of SN-GRB: SNe Ic-bl with GRBs }\label{sngrb_sec}

While the explanation for GRBs after their initial discovery included a vast array of different theories, intensive follow-up observations of GRBs over the last two decades have established that long-duration soft-spectra GRBs \citep{kouveliotou93}, or at least a significant fraction of them, are directly connected with supernovae and result from the cataclysmic death of massive, stripped stars (see review by \citealt{woosley06_rev}). The most direct proof of the SN-GRB association comes from spectra taken of the GRB afterglows, where the spectral fingerprint of a SN, specifically that of a broad-lined SN Ic, emerges over time in the spectrum of the GRB afterglow. Near maximum light, GRB-SNe appear to show broad absorption lines of O I, Ca II, and Fe II (see Fig.~\ref{SNclass_fig}), while there is no photospheric spectrum of a confirmed GRB-SN that indicated the presence of H or showed optical lines of He I (see also below). 

Below we briefly list the SN-GRB cases, in descending order of quality and quantity of data (see also Table 1 in \citealt{woosley06_rev} and detailed discussions in \citealt{hjorth11}). 
The five most solid cases of the SN-GRB connection, with high signal-to-noise and multiple spectra, are usually at low $z$ :
SN1998bw/GRB980425  at $z=0.0085$ \citep{galama98}, SN2003dh/GRB030329 at $z=0.1685$     
\citep{stanek03,hjorth03,matheson03}, SN2003lw/GRB031203 at $z=0.10058$ \citep{malesani04}, SN2006aj/GRB060218 at $z=0.0335$ \citep{campana06,modjaz06, pian06,sollerman06,mirabal06,cobb06,kocevski07}, and most recently, SN2010bh/GRB100316D at $z=0.0593$ \citep{chornock10,starling11}, where the SN spectra lines were visible as early as 2 days after the GRB \citep{chornock10}. Two special SNe, SNe 2008D and 2009bb, and the potential presence of a jet in them will be discussed below.


Again, it is important to note that the spectra of the observed GRB-SNe are not those of any kind of core-collapse SNe, but specifically those of SN Ic-bl. The fact that there is no longer the large H envelope present when the star explodes as a SN-GRB is a crucial aspect of why and how the jet can drill its way trough the star \citep{woosley93,zhang04}. In addition, SN-GRB do not show the optical Helium lines in their spectra.  While there is some discussion of He in the spectrum of SN 1998bw/GRB 980425 \citep{patat01}, its claim is based on a broad spectral feature at 1 micron, which could be due to lines other than He I $\lambda$ 10830  \citep{millard99,gerardy04,sauer06}. The NIR spectrum of the most recent  SN2010bh/GRB100316D, did not show the 1 micron He line \citep{chornock10}. What's more, it remains note-worthy and peculiar that almost all of the solid SN-GRB connections are with GRBs that are usually regarded as non-classical: i.e., GRBs that  are less beamed (opening angles with > 30$-$80 deg), have low gamma-ray luminosity (i.e., $L_{\gamma}^{iso}  \le 10^{49}$ \ergs), soft spectra, and thus, are also sometimes called X-ray Flashes (XRFs) or X-ray rich GRBs, being perhaps more common than cosmological GRBs \citep{cobb06,soderberg06_06aj,guetta07}. Only GRB~030329 connected with SN~2003dh is a classical GRB whose kin we see at high-$z$. Either those cosmological high-luminosity GRs are rare at low-$z$, where we can see the SN signatures spectroscopically, or the SN-GRB connection is confined to only GRBs that are more isotropic and of low luminosity. For reference, a SN with the same large luminosity as \snbw\ will appear at $R\sim$22 mag at $z$=0.5, so approaching the limit of obtaining a spectrum with a large-aperture telescope and reasonable exposure times. 

The second broad class encompasses cases with only one epoch of low S/N spectra, which are at higher $z$:  XRF~020903 at $z=0.25$ \citep{soderberg05,bersier06}, SN~2002lt/GRB~021211 at $z=1.006$ \citep{dellavalle03},  SN~2005nc/GRB~050525A at $z=0.606$ \citep{dellavalle06_05nc}. The following last class of possible SN-GRB connections is based on observed rebrightening in the light curves of GRB afterglows that are consistent with emerging SN light curves (e.g., \citealt{bloom99}), and in some cases with multi-color light curves that constrain the SED. While a a few high-$z$ cases have high-quality data that make them convincing  (e.g., most recently \citealt{cobb10,cano11}), there are many more where the data is of lower quality (e.g., \citealt{zeh04}), making the SN interpretation in all cases less secure. The earliest, but more indirect, hints for the existence of a link between GRBs and the death of massive stars was the detection of star-formation features in the host galaxies of GRBs \citep{djorgovski98,fruchter99} and correlation of GRB positions in their host galaxies with starforming regions \citep{bloom02}.

While two GRBs, GRBs 060505 and 060614, have been observed without a bright SN (\citealt{dellavalle06,fynbo06,gal-yam06}), it is debated whether those were indeed bona fide long-GRBs \citep{gehrels06,zhang07,bloom08}, posing another challenge to the GRB classification scheme \citep{bloom08}. In any case, it is fair to say that for any bona fide and un-ambiguous long-GRB with a sufficiently low redshift to enable a spectroscopic SN detection, a broad-lined SN  Ic has been detected. This does not apply to X-ray flashes, where multiple searches for SN signatures in low-$z$ XRFs have not yielded clear evidence for associated SNe \citep{levan05,soderberg05}.

\section{Do all SNe Ic-bl have an accompanied GRB? }\label{offaxis}

While the list of SN-GRB connections is short, there is a growing number of SN Ic-bl that are discovered by various SN surveys\footnote{For a full list of IAUC-announced SNe, see the following link: http://www.cfa.harvard.edu/iau/lists/RecentSupernovae.html }, which are not observed to have an accompanied GRB or to be engine-driven (except SN Ic-bl 2009bb, see below). One plausible explanation may be that all SN Ic-bl have an accompanied GRB, but are not observed by us due to viewing-angle effects: our line-of-sight may not intersect the collimated jet of GRB emission and thus, we may not detect gamma rays for these so-called "off-axis" GRBs. Various investigations have been trying to address this viewing angle effect and to find off-axis GRBs.

\subsection{Search for Off-Axis GRBs}

Even for the most highly-beamed GRB, the GRB jet gets decelerated over time and becomes effectively an isotropic blast wave such that the jet that was initially beamed away from our line of sight produces afterglow emission (so-called "orphan afterglows") which we may see to increase over time scales from months to several years (e.g., \citealt{perna98,vaneerten10}). If GRB jets are highly beamed, off-axis GRBs and their subsequent orphan afterglow are a natural prediction. Thus, various wide-field searches have looked for them in the optical and radio wavelengths (e.g., \citealt{levinson02,malacrino07}), but none has been detected at high significance. Along the same vein, specifically \citet{soderberg06_radioobs} and \citet{soderberg10_09bb} targeted as part of their survey 143 SN Ib/c (including SN Ic-bl) for late-time VLA observations to search for off-axis GRB afterglows, but none of their objects showed evidence for bright, late-time radio emission that could be attributed to off-axis jets coming into our line of sight, until 2009. In 2009, SN~Ic-bl 2009bb was discovered optically and exhibited a large radio luminosity \citep{soderberg10_09bb,pignata11} that requires substantial relativistic outflow with $10^3$ more matter coupled to relativistic ejecta than expected from normal core-collapse, and thus, arguing for an engine-driven SN. While no coincident GRB was detected, it is not clear whether it is  because there was a weak GRB that went undetected or the GRB was off-axis or there were no gamma-rays produced during the SN. While also SN~Ic~2007gr was claimed to indicate an engine-driven explosion without an observed GRB \citep{paragi10}, its radio light curves and X-ray data indicate that it may well be an ordinary SN Ib/c explosion \citep{soderberg10_07gr}.   

\subsection{Relative Rates of SN Ic-bl vs LGRB}

A statistical approach for understanding the SN-GRB connection is to compare the explosion rates of broad-lined SN Ic to those of Long GRBs and see if they are comparable. While this line of argument is very reasonable, it is not very conclusive at this point, given that both kinds of rates are somewhat uncertain. On the GRB side, the beaming angle is uncertain, given the possible 2 different populations of GRBs (high-luminosity, highly-beamed vs. low-luminosity and nearly isotropic)  - on the SN side, the rates of SN subtypes are not well known, specifically those of SN Ic-bl \citep{li11}, as well as how selection effects may enter differently for GRB and SN searches \citep{woosley06_rev}. Nevertheless, \citet{guetta07} investigated this question by distinguishing between High- and Low-luminosity GRBs and by deriving the SN Ic-bl rate from a heterogenous list of SNe discovered by different surveys. They estimate that the ratio of low-luminosity GRBs to SN Ic-bl is in the range of  $\sim$1\%$-$10\%, assuming that  SN Ic-bl live in the same environments as SN-GRBs and have the same host galaxy luminosity, $M_B$, which is at odds with recent observations (Section~\ref{metallicity_sec}). Independently, the extensive radio search for off-axis GRBs in 143 optically discovered SN Ib/c (not strictly only SN Ic-bl) yields that less than $\sim$1\% of SN Ib/c harbor central engines \citep{soderberg10_09bb}, thus broadly consistent with the above estimates, depending on the SN Ic-bl rate. 

In conclusion, it appears that SN-GRB are intrinsically rare and that certain conditions must be fulfilled for an exploding, massive and stripped star to simultaneously produce a GRB jet and to release a large amount of energy.

\section{Aspherical Explosions: Only in SN-GRBs?}\label{asphericity_sec}

One of the fundamental questions in the SN-GRB field is whether aspherical explosions are the exclusive and distinguishing property of GRB-SN, or whether they are generic to the core-collapse process. Besides polarization measurements, late-time spectroscopy is a premier observational tool for studying the geometry of the SN explosion.
At late times ($> 3-6$ months), the whole SN ejecta become optically thin in the continuum and hence affords a deeper view into the core of the explosion than spectra taken during the early photospheric phase. Moreover, the emission line shapes provide information about the velocity distribution of the ejecta \citep{fransson87,schlegel89}, and thus its radial extent, since the ejecta are in homologous expansion (where $v_r  \propto\ r$). A radially expanding spherical
shell of gas produces a square-topped profile, while a filled uniform
sphere produces a parabolic profile. In contrast, a cylindrical ring,
or torus, that expands in the equatorial plane gives rise to a
``double-peaked'' profile as there is very little low-velocity
emission in the system, while the bulk of the emitting gas is located
at $\pm v_t$, where $v_t$ is the projected expansion velocity at the
torus.

\begin{figure}[!ht]
  \includegraphics[height=.5\textheight]{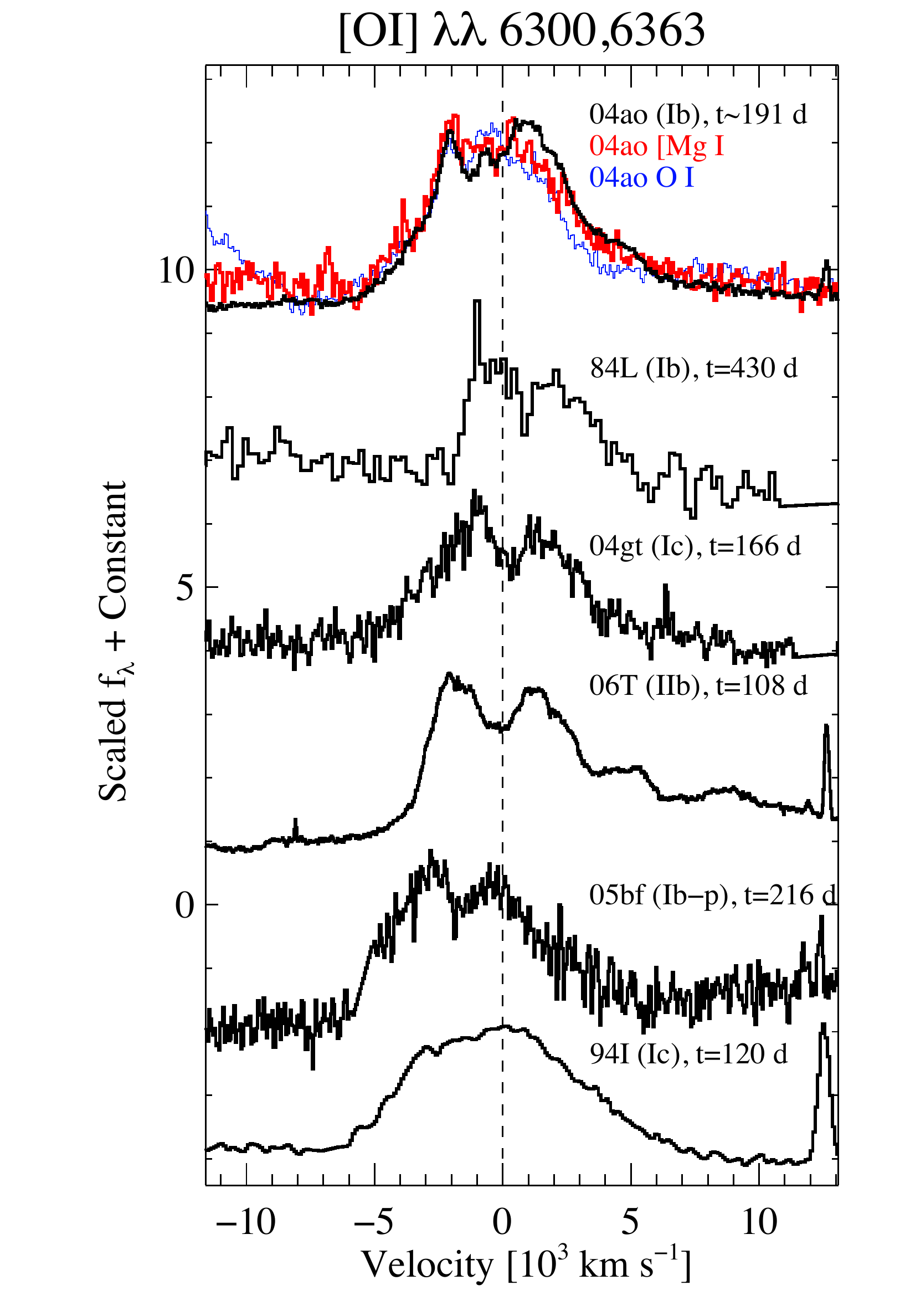}  
  \caption{Montage of five SNe with double-peaked oxygen profiles in velocity
space. SN name, type, and phase of spectrum (with respect to maximum light) are indicated. SN~1994I \citep{filippenko95}, which exhibits a single-peak oxygen line profile, is plotted
for comparison at the bottom. The dashed line marks zero velocity with respect
to 6300 $\lambda$. For SN 2004ao, we plot the scaled profiles of O I $\lambda$7774 (in blue)
and [Mg I $\lambda$4571 (in red), which are not doublets, but also exhibit the two
peaks. As discussed in the text, the two horns are unlikely to be due
to the doublet nature of [O I] $\lambda\lambda$6300, 6364. From \citet{modjaz08_doubleoxy}. }
\label{OIdouble_fig}
\end{figure}

Thus, a number of studies embarked on the difficult undertaking of obtaining such nebular spectra with adequate resolution and signal for a number of stripped SNe. As the objects are usually faint at that stage (19$-$23 mag), they call for large-aperture telescopes. First, \citet{mazzali05_03jd} and then \citet{maeda07} reported that SN Ic-bl 2003jd and the peculiar SN Ib 2005bf, respectively, displayed  a double-peaked profile of [O I]$\lambda\lambda$6300,6364 in nebular spectra. Subsequently, \citet{modjaz08_doubleoxy} and \citet{maeda08} independently presented a large number of stripped SNe displaying pronounced double-peaked profiles of [O I]$\lambda\lambda$6300,6364,  the strongest line in late-time spectra of SNe without H, with velocity separations ranging between 2000 to 4000 \kms\ (see Fig.~\ref{OIdouble_fig}). Those profiles were interpreted as indicating an aspherical distribution of oxygen, possibly in a torus or flattened disc seen edge-on, suggesting that strong asphericity is ubiquitous in core-collapse SNe, and not necessarily a signature of an association with
a GRB. For SN spectra with sufficient S/N in some of the additional lines ( \OxySeven, \MgOne\ ) and with multiple epochs (e.g., SNe 2004ao, 2008D, \citealt{modjaz09,tanaka09_08dlate}), the double-peaked profiles are unlikely to be caused by known optical depth effects. Furthermore, \citet{taubenberger09} presented and analyzed a large set of 98 late-time spectra of a total of stripped SNe (some of which were taken from the literature) and found a rich phenomenology of line structures. The results of their statistical analysis suggest that probably at least half of all stripped SNe are aspherical and that line profiles are indeed
determined by the ejecta geometry, with Mg and O similarly distributed within the SN ejecta.

Recently, \citet{milisavljevic10} presented high S/N and multi-epoch nebular spectra of a select number of stripped SNe (including published data) and, coupled with detailed spectral line analysis and fitting, raised important questions about the interpretation of the indicated geometry. They suggested that alternative geometries beyond torus or highly flattened disks are possible for some of the SNe, where the double-peaked oxygen profile could be either coming only from preferentially blueshifted emission with internal obscuration in the red, or could consist of two separate emission components (a broad emission source centered around zero velocity and a narrow, blue-shifted source). Future high-S/N, multi-epoch and multiple-line observations of a large sample of SNe II/IIb/Ib/Ic coupled with radiative transfer models should help to elucidate the observed blue- and redshifts of the line profiles and constrain the exact geometry.

In conclusions, observed double-peaked oxygen lines are not necessarily a proxy of a mis-directed GRB jet and they suggest that asphericities (of whatever exact geometry they may be) are most likely prevalent in normal core-collapse
events. This result that asphericites are an ubiquitous feature during core-collapse  is in line with similar
conclusions based on polarization studies of SN II, Ib and SN Ic
e.g., \citep{leonard05,maund09_08d}, neutron-star kick velocities \citep{wang06_nskicks}, 
young SN remnant morphologies \citep{fesen06_casabi}, and theoretical
modeling efforts \citep{scheck06,burrows06,dessart08}. Aspherical explosion geometry does not appear to be distinguishing feature of SN-GRBs, though SN-GRBs may have the highest degree of asphericity according to some models \citep{maeda08}.

\section{Progenitor Mass as the Culprit?}\label{mass_sec}

One obvious possibility is that progenitor mass, one of the most fundamental properties of a star, may set apart SN with GRBs from those without GRBs. Specifically, the SN-GRB progenitors could be of high mass (enough to produce a BH required for the collapsar model) and higher than the progenitors of SN without GRBs. In order to estimate the mass of the SN-GRB progenitors, one has to use a different method than the direct pre-explosion imaging technique (which even with HST's exquisite resolution can only be used for SN progenitor searches for up to $\sim$20 Mpc), since the GRB and SN progenitors are at much larger, cosmological distances and since even in the local universe, detection attempts of stripped SN progenitors have failed \citep{smartt09}. There are two different techniques for indirectly estimating the main sequence (MS) mass of a SN/GRB progenitor. The first one consists of modeling the spectra and light curves of the individual SN/GRB in order to constrain the ejecta mass and core-mass before explosion and then use stellar evolutionary codes (e.g., see Fig. 1 in \citealt{tanaka09_08dearly}) to infer the MS mass, subject to the caveats of uncertain mass loss rates and rotation. The second technique entails studying the stellar population at the SN/GRB position as a proxy for the SN/GRB progenitor. 

The first technique has been performed for a small number of SN/GRBs (e.g., \citealt{mazzali06_03lw}, for a review see \citealt{tanaka09_08dearly,nomoto10}), mostly for the nearby GRB-SN and a few peculiar SN and its results suggest that the SN-GRB are from the more massive end of stellar masses ($\sim$ 20$-$50 \msol), but not necessarily from the most massive stars. However, so far only data of two SNe Ic-bl without an observed GRB (SNe 1997ef and 2002ap) have been modeled, and thus, it is not clear from this line of research whether the GRB-less SN Ic-bl progenitors are as massive as those of SN-GRB.

On the other hand, the second technique of studying the stellar populations at the explosion sites has been performed on a statistical set of different types of SN, SN-GRB and GRBs (those that are at higher redshifts)  by comparing the amount of light at the position of the GRB or SN (after it had faded) to that of the rest of the host galaxy, as a proxy for the amount of star formation. \citet{fruchter06} and \citet{svensson10} found that GRBs are more concentrated towards the brightest regions of their host galaxies than are SN II (for the same range of high-$z$), and took their data to indicate larger progenitor masses for GRBs than for SN II, which is consistent with SN II pre-explosion detections that indicate modest MS-progenitor masses of 8$-$16 \msol\ for SN IIP (see \citealt{smartt09} for a review). 
Importantly, \citet{kelly08} demonstrate, using the same technique as \citet{fruchter06}, that nearby ($z<0.06$) SNe Ic are also highly concentrated on the brightest regions within their host galaxies, thus
implying similarly high progenitor masses for SNe Ic without GRBs, as for
GRBs themselves. Thus, these observations suggest another ingredient for GRB production besides
higher mass progenitors. While \citet{anderson08,anderson09} have similar findings, their interpretation differs, as they regard  the increased centralization of a SN distribution to imply increased progenitor metallicity, not increased progenitor mass. We will turn to the question of metallicity  in the next section.

\section{Metallicity as the Culprit?  }\label{metallicity_sec}

Metallicity is expected to influence not only the lives of massive
  stars but also the outcome of their deaths as supernovae (SNe) and
  as gamma-ray bursts (GRBs). However, before 2008, there were surprisingly few
  direct measurements of the local metallicities of SN-GRBs, and virtually none for the various types of core-collapse SNe.
  
Before delving into the details of the metallicity studies, let us explain what we refer to when using the term ÒmetallicityÓ. Theorists usually refer to the iron mass fraction of the SN progenitor, which is important for setting the mass loss rate of the pre-explosion massive star, since the bulk of the opacity is provided by iron and its huge number of lines (\citealt{vink05}). Observers, on the other hand, usually measure the oxygen abundance of HII regions  of some (usually central) part of the host galaxy or, in the best-case-scenario, that at the SN positions\footnote{We also note that when we discuss oxygen, we do not refer to the oxygen that was released during explosion (Section~\ref{asphericity_sec}), since it usually takes $10^5-10^7$ years of settling time for the SN yields to be incorporated into the ISM, and we are observing the environments only months to years after explosion. While there may be concerns about "self-enrichment", i.e., by evolved stars in HII region before explosion (such that the measurements would not reflect the natal metallicity but some self-polluted, higher value), many HII regions do not show clear signs of self-enrichment  \citep{wofford09}.}. The nebular oxygen abundance is the canonical choice of metallicity indicator for ISM studies, since oxygen is the most abundant metal, only weakly depleted onto dust grains (in contrast to refractory elements such as Mg, Si, Fe, with e.g., Fe being depleted by more than a factor of 10 in Orion; see \citealt{simondiaz11-orion}), and exhibits very strong nebular lines in the optical wavelength range (e.g., \citealt{pagel79,osterbrock89,tremonti04}). Thus, well-established diagnostic techniques have been developed (e.g., \citealt{kewley02,pettini04,kobulnicky04,kewley08}). Due to the short lifetimes of the massive SN and GRB progenitor stars ($\le$ 10 million years for 20 \sm\ star, \citealt{woosley02}), we do not expect them to move far from their birth HII region sites (but see  \citealt{hammer06,eldridge11} and below) and thus, we take the abundance of the HII region at the SN site to indicate the natal metallicty of the SN or GRB progenitor. In one GRB case, where there is an independent metallicity measurement from absorption-line ratios in the Xray spectra from the circumburst medium of \snaj\ \citep{campana08} and the common nebular oxygen-abundance measurement (e.g., \citealt{modjaz06}), the two completely independently derived values are in broad agreement. Furthermore, it appears that gas-phase oxygen abundances track the abundances of massive stars well, as seen in a number of studies for the Orion nebula (see \citealt{simondiaz11-orion} for a good review) and for blue supergiants in NGC 300 \citep{bresolin09}.

When considering oxygen abundance measurements, one has to remember the long-standing debate about which diagnostic to use, as there are systematic metallicity offsets between different methods (recombination lines vs. collisionally excited lines vs. "direct" method) and different strong-line diagnostics (see \citealt{kewley08} and \citealt{moustakas10} for detailed discussions), as well as the debate about the solar oxygen abundance value \citep{asplund09_rev}. Nevertheless, the (relative) metallicity trends can be considered robust, if the analysis is performed self-consistently in the same scale, and trends are seen across different scales. We demonstrate the power and potential of this approach in the next subchapters.

\subsection{Metallicity of SNe with and without GRBs}
\begin{figure*}[!ht]
 \includegraphics[scale=0.55,angle=-90]{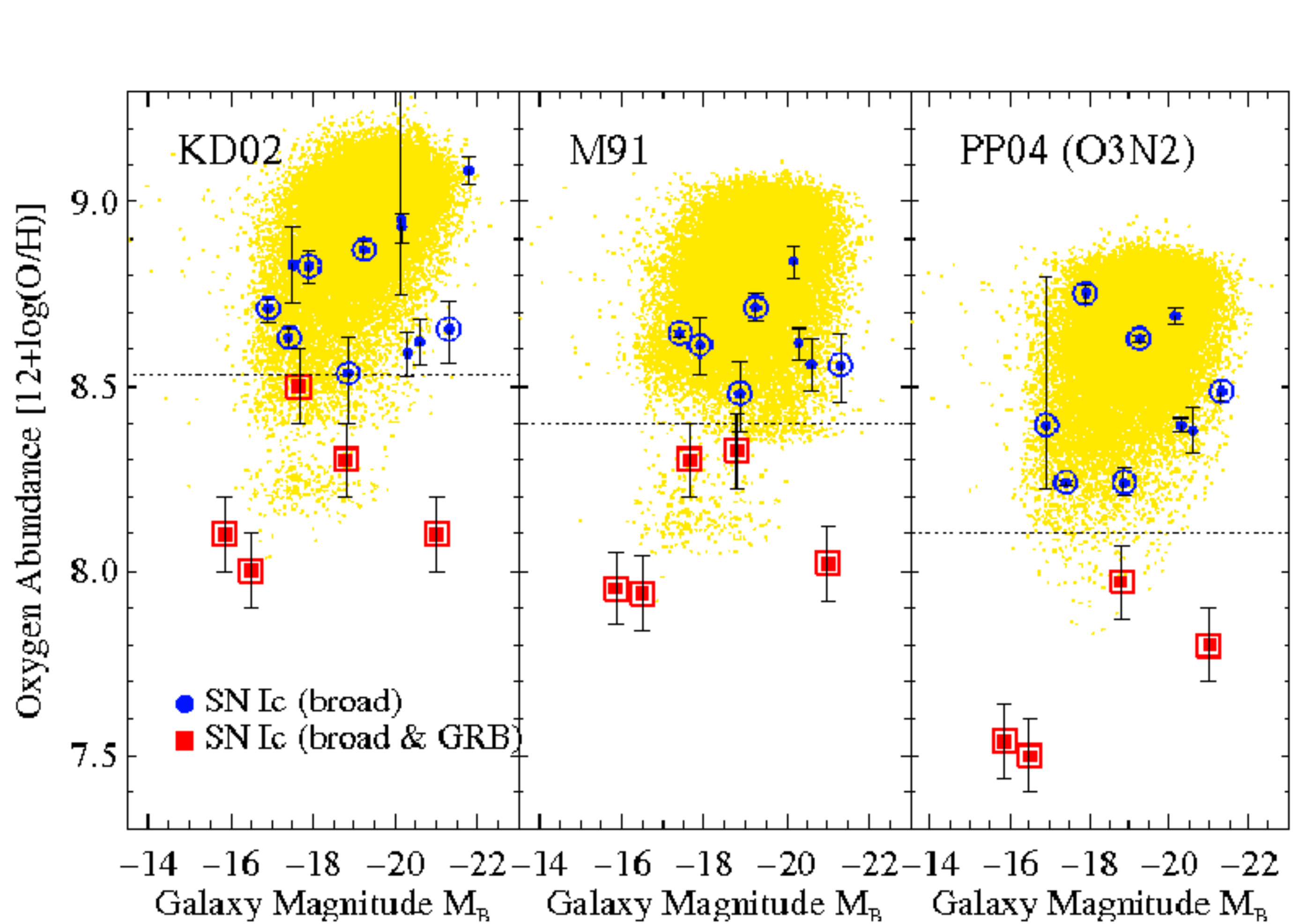} 

  \caption{Host-galaxy luminosity
    ($M_B$) and host-galaxy metallicity (in terms of oxygen abundance)
    at the sites of nearby ordinary broad-lined SNe Ic (SN Ic (broad):
    blue filled circles) and broad-lined SNe Ic connected with GRBs
    (SN (broad \& GRB): red filled squares) in three different,
    independent metallicity scales. Host environments of
      SNe-GRBs are more metal poor than host environments of
      broad-lined SNe Ic where no GRB was observed, for a similar
      range of host-galaxy luminosities and independent of the
      abundance scale used. For reference, the yellow points are
    nuclear values for local star-forming galaxies in SDSS (Tremonti
    et al 2004), re-calculated in the respective metallicity scales,
    and illustrate the empirical luminosity-metallicity relationship
    for galaxies. The host environment of the most recent SN-GRB \citep{chornock10,starling11} is consistent with this trend. From \citet{modjaz08_Z}.}
\label{oxy-modj08-fig}
\end{figure*}

Many theoretical GRB models favor rapidly rotating
massive stars at low metallicity \citep{hirschi05, yoon05,woosley06_z,langer06}
as likely progenitors. Low metallicity seems to be a promising
route for some stars to avoid losing angular momentum from
mass loss \citep{vink05,crowther06} if the mass loss mode is set by line-driven, and therefore, metallicity-driven winds (but see \citealt{smith06} for eruptions in some massive stars). If the stellar core is coupled to the outer envelopes via, e.g., magnetic torques \citep{spruit02}, it is able to retain its high angular momentum preferentially at low metallicity. In turn, high angular momentum in the core appears to be a key ingredient
for producing a GRB jet for both the collapsar and the magnetar models.

If the GRB progenitor is supposed to be at low metallicity for minimal mass loss, then how does it remove its outer layers, especially the large Hydrogen envelope, for we do not see any trace of H or He in the optical spectra of SN-GRB? Either via binaries (e.g., \citealt{fryer07,podsiadlowski10}) or, if the abundances in the star are sufficiently low, and thus, the star rotates rapidly enough, then quasi-chemical homogeneous evolution may set in, where hydrogen gets mix into the burning zones of the star via rotational mixing \citep{maeder87,langer92,heger00,maeder00_rev}, such that the star has low hydrogen abundance and a large core mass just before explosion. This mechanism seems plausible for producing a GRB and a broad-lined SN Ic at the same time, though it is debated whether it can explain all observed trends in the VLT FLAMES survey of massive stars at different metallicities \citep{hunter09,frischknecht10,brott11}. 

Before 2007, a number of studies showed observationally that GRB hosts are of lower luminosity compared to core-collapse SN hosts \citep{fruchter06,wolf07} and, when measurable, of low metallicity (e.g., \citealt{fynbo03,prochaska04,sollerman05,modjaz06}), especially compared to the vast majority of SDSS galaxies \citep{stanek07}. The next step was to compare the abundances of SNe Ic-bl with GRBs to SNe Ic-bl intrinsically without GRBs to test whether low metallicity is a necessary condition for GRB production. In 2007 and 2008 we embarked on directly measuring metallicities of a statistically significant
sample of broad-lined SN Ic environments and deriving them in the same
fashion, which we presented in our study of \citet{modjaz08_Z}, the first of its kind. There, we compared the chemical abundances at the sites of 5 nearby ($z <$ 0.25) broad-lined SN Ic that accompany nearby GRBs with those of 12 nearby ($z <$ 0.14) broad-lined SN Ic that have no observed GRBs.  We showed that the oxygen abundances at the GRB sites are systematically lower than those found at the sites of ordinary broad-lined SN Ic (Fig.~\ref{oxy-modj08-fig}). Unique features of our analysis included presenting new spectra of the host galaxies and analyzing the measurements of both samples in the same set of ways, via three independent metallicity diagnostics, namely those of \citet{kewley02} (KD02), \citet{mcgaugh91} (M91) and \citet{pettini04} (PP04). We demonstrated that neither SN selection effects (SN found via targeted vs. non-targeted surveys, for an extensive discussion see Section~\ref{future_sec}) nor the choice of strong-line metallicity diagnostic could cause the observed trend. Though our sample size was small, the observations (before 2009) were consistent with the hypothesis that low metal abundance is the cause of some massive stars becoming SN-GRB. While each metallicity diagnostic has its own short-coming, if we use the scale of PP04, which has been suggested by \citet{bresolin09} to be the strong-line method in most agreement with abundances from stars, then the "cut-off" metallicity value would be $\sim$ 0.3 \sz. Furthermore, a comparison between the local metallicity of the GRB-SN site and the global host galaxy value via resolved metallicity maps yields that the GRB-SN local values track the global host value, but are also amongst the most metal-poor site of the galaxy \citep{christensen08,levesque11_10bh}.

This was in 2008 - \textbf{now}, however, the case for a metallicity threshold is much less clear. Over the last three years, two "dark" GRBs  \citep{graham09,levesque10_darkgrb} and one radio-relativistic SN, SN 2009bb \citep{soderberg10_09bb,levesque10_09bb} have been observed at high, super-solar, metallicity. Even if one includes those higher metallicity explosions regardless of whether they share the same progenitor channels as SN-GRBs, \citet{levesque10_grbhosts} show that the M-Z relationship for GRBs lies systematically below that of the bulk of the normal starforming galaxies in the corresponding redshift ranges (see their Fig. 1). While there are suggestions that the observed low-metallicity trend could be partly produced by the newly-discovered relationship between host galaxy metallicity, mass and star formation rate (so-called "fundamental" metallicity relation, \citealt{mannucci10}) such that low metallicity galaxies have high star formation rates \citep{mannucci11,kocevski11}, it does not explain why there are not more GRBs in intermediate- and high-mass galaxies \citep{kocevski09,kocevski11} and the observed evolution in the GRB rate density with increasing redshifts \citep{butler10}. It is currently hotly debated whether dust may be the reason for the metallicity offset (e.g., \citealt{fynbo09}), since it could obscure the optical afterglows of GRBs at high metallicity, thus explaining the lack of high-mass and high-metallicity GRB host galaxies (since usually optical afterglows are needed for precise localization of the host).

An obvious test will be to construct the $M-Z$ relationship for other explosions that track massive star formation, such as normal SN Ib and SN Ic, found in the same fashion as GRBs, namely in non-targeted surveys such as the Palomar Transient Factory (see Section~\ref{future_sec}), and to compare it to that of GRBs. 

\subsection{Metallicity of various regular types of CCSN}\label{ccsnz_sec}

For illuminating the SN-GRB connection and for pursuing stellar forensics on the specific SNe Ic-bl with and without GRBs, it is also important to gain an understanding of the progenitors of "normal" stripped SNe. Here too, the two outstanding progenitor channels are  either
single massive Wolf-Rayet (WR) stars with main-sequence (MS) masses of
$\ga$ 30 \sm\ that have experienced mass loss during the MS and WR
stages (e.g., \citealt{woosley93}), or binaries from lower-mass
He stars that have been stripped of their outer envelopes through
interaction (\citealt{podsiadlowski04}, and references therein), or a
combination of both. Attempts to directly identify SN~Ib/c progenitors
in pre-explosion images have not yet been successful
(e.g., \citealt{gal-yam05,maund05}; \citealt{smartt09_rev}). 

A more indirect but very powerful approach is to study the
environments of a large sample of CCSNe in order to discern any systematic
trends that characterize their stellar populations. Discussed already was the study of the amount of blue light at the position of different types of explosions (Section~\ref{mass_sec}), which indicates that stripped SNe, and especially SNe Ic, are more concentrated towards the brightest regions of their host galaxies than SN II  \citep{kelly08,anderson08}, possibly suggesting
the progenitors of SNe~Ib/c may thus be more massive than those of
SNe~II, which are $\sim 8$--16 \sm\ (see \citealt{smartt09_rev} for a
review). 

There exist a few metallicity studies of CCSN host environment 
that either \textbf{indirectly} probe the metallicities of a \textbf{large} set of SNe II, Ib and Ic or that \textbf{directly} probe the local metallicity of a \textbf{small} and select set of interesting/peculiar stripped SNe. Nevertheless, interesting trends have emerged:
Studies to measure the metallicity by 
using the SN host-galaxy luminosity as a proxy
\citep{prantzos03,arcavi10}, or by using the metallicity of the galaxy
center measured from Sloan Digital Sky Survey (SDSS) spectra
\citep{prieto08} to extrapolate to that at the SN position
\citep{boissier09} find a) that host galaxies of SNe Ib/c found in targeted surveys seem to be in more luminous and more metal-rich galaxies than those of SNe II \citep{prieto08,boissier09} and b) that SNe Ic are missing in low-luminosity and presumably, low-metallicity galaxies of the untargetted survey PTF, while SN II, Ib, and Ic-bl are abundant there \citep{arcavi10}.  Those prior metallicity studies do not directly probe the local environment of each SN (which can be different from the galaxy
center due to metallicity gradients) nor do some differentiate between
the different SN subtypes.
In \citet{modjaz11}, we presented the largest existing set of
  host-galaxy spectra with H~II region emission lines at the sites of
35 stripped-envelope core-collapse SNe and including those from the literature and from \citet{modjaz08_Z}, we analyzed the metallicity environments of a total of 47 stripped SNe. We derived local oxygen abundances in a robust manner in order to constrain the
  SN~Ib/c progenitor population. We obtained spectra at the SN sites,
  included SNe from targeted and untargeted surveys, and performed
  the abundance determinations using the same three different oxygen-abundance
  calibrations as in \citet{modjaz08_Z}. We found that the sites of SNe~Ic (the demise of the most
  heavily stripped stars having lost both H and He layers) are
  systematically more metal-rich than those of SNe~Ib
  (arising from stars that retained their He layer) in all calibrations. A
  Kolmogorov-Smirnov-test yielded the low probability of 1\% that SN~Ib and SN~Ic environment
  abundances, which are different on average by $\sim$0.2 dex (in the Pettini \& Pagel scale), are drawn from the same parent population. Broad-lined SNe~Ic (without GRBs) occur at metallicities between those of SNe~Ib
  and SNe~Ic. Lastly, we found that the host-galaxy central oxygen
  abundance is not a good indicator of the local SN metallicity (introducing differences up to 0.24 dex), and concluded that large-scale SN surveys need to obtain local abundance measurements in order to quantify the impact of metallicity on stellar death. 
  
  A reasonable suggestion for why the environments of SNe~Ic are more
metal rich than those of SNe~Ib is that metallicity-driven winds
\citep{vink05,crowther06} in the progenitor stars prior to explosion
are responsible for removing most, if not all, of the He layer whose
spectroscopic nondetection distinguishes SNe~Ic from SNe~Ib. This
explanation may favor the single massive WR progenitor scenario as the
dominant mechanism for producing SNe~Ib/c \citep{woosley93}, at
least for those in large star-forming regions. While the binary scenario has been suggested
as the dominant channel for numerous reasons (see
\citealt{smartt09_rev} for a review; \citealt{smith11_snfrac}), we
cannot assess it in detail, since none of the theoretical studies
(e.g., \citealt{eldridge08}, and references therein) predict the
metallicity dependence of the subtype of stripped SN.  However, our
results are consistent with the suggestion of \citet{smith11_snfrac}
that SNe~Ic may come from stars with higher metallicities (and masses)
than SNe~Ib, even if they are in binaries. Furthermore, the finding that the metallicity environments for SN Ic-bl are different from those of SN Ic indicates that their progenitors may be physically different (perhaps because of magnetic fields or other factors) and that the presumably significant amount of mass at high velocities in SN Ic-bl are probably not only due to viewing-angle effects.
    
  Most recently, similar studies for SN Ib and SN Ic have been conducted by \citet{anderson10} and \citet{leloudas11}. While they do find small differences between SN Ib and SN Ic, with SN Ic in slightly more metal-rich environments, they conclude that their findings are not statically significant. While the reasons for these different
metallicity findings in different studies are not yet clear, some of the aspects of their studies may complicate direct SN~Ib vs. SN~Ic metallicity comparisons with statistical power.  For example, historical SNe~Ib/c
without firm subtype classifications (e.g., SNe 1962L, 1964L) from
only targeted surveys, some with incorrect SN offsets as announced in
the IAUC (e.g., for SNe 1987M and 2002ji; S. Van Dyk 2010, private
communication) are included \citep{anderson10} and an unequal number of SN Ib (N=14) and SN Ic (N=5) in \citet{leloudas11}. In any case, definite answers should be provided by future environmental metallicities studies using a very large SN crops from the same, homogeneous and galaxy-unbiased survey, such as the one we are undertaking (see Section~\ref{future_sec}), which should be ideally suited to determine the environmental conditions that influence the various kinds of massive stellar deaths in an unbiased fashion.

  
 \subsection{SN and GRB Host Metallicity Measurements as a Rapidly Expanding Field }\label{expandingfield_sec}

Not only for stripped SNe and SN-GRBs, but also for other kinds of SNe and transients have metallicity studies emerged as a promising tool to probe their progenitor and explosion conditions. Another class of CCSN that has piqued a lot of interest in the past few years is the emerging field of over-luminous SNe, i.e., SNe defined as more luminous in absolute magnitudes than $M_V\sim-21$ mag \citep{smith07_06gy,ofek07_06gy,quimby11}, that are being discovered in wide-field surveys. Is is hotly debated what powers their optical brilliance, whether it's due to circumstellar interaction, large amount of synthesized \synNi\ during the explosion of a pair-instability SN or the birth of a magnetar (\citealt{kasen10}). Host galaxy studies show that 
their host galaxies are of low luminosity, highly starforming and blue \citep{neill11}, similar to GRB-host galaxies, and similarly of low-metallicity, when measured \citep{stoll11}, with the exception of the host of SN~2006gy, the SN which was first claimed as a pair-instability SN \citep{smith07_06gy,ofek07_06gy}. Furthermore, the best candidate for a pair-instability SN, SN~2007bi \citep{gal-yam09} has a host galaxy with a metallicity of 12+log(O/H)$_{M91}=8.15 \pm 0.15$ in the McGaugh scale, and thus, $\sim0.3$ \sz\ \citep{young10}, so it is a subsolar galaxy, but not of extreme subsolar metallicity, as one might expect from Pop III stars in the high-$z$ universe. Thus, if SN~2007bi is representative of pair-instability SNe, then they should be found frequently during current and next generation of wide-area surveys, which have enough volume to discover rare transients.

Furthermore, even for SN Ia, which arise from the thermonuclear explosion of a white dwarf at or near the Chandraskhar mass limit in a binary system, host galaxy studies have uncovered trends for SN Ia luminosity with host galaxy morphology (e.g., \citealt{hamuy96}) and mass (e.g., \citealt{kelly10,sullivan10}), where more luminous SN Ia tend to be in more luminous and (assuming the luminosity-metallicity relationship for galaxies) metal-rich galaxies, which is consistent with measured metallicity studies \citep{gallagher08}. However, for SN Ia with their long delay times (200 Million yrs to a few Gigayears, e.g., \citealt{maoz10}) and the associated large offsets between birth and explosion sites, it is not clear whether measuring the gas-phase metallicity (which reflects that of the currently starforming gas) at the SN position really reflects the natal metallicity of the old progenitor \citep{bravo11}. Nevertheless, integrated metallicities from stars in the host galaxy \citep{gallagher08} or those of dwarf galaxies \citep{childress11}, which usually have a small spread in metallicities, may be revealing.


\section{SN 2008D/XRT080109: Stellar Forensics by Witnessing the Death Throes of a Stripped Star}\label{08d_sec}

A complimentary stellar forensics tool is to catch the massive star during its death throes. This happened for SN 2008D/XRT080109 (where XRT stands for X-ray Transient) which was discovered by \citet{soderberg08}. From the early light of the explosion, one can reconstruct a massive star's pre-explosion composition and radius, which provides a powerful tool to closely investigate a single star out to cosmological distances. Besides its utility, the story of \snxray\ reminds us of the importance of serendipity in science. \snxray\ was discovered by \citet{soderberg08} in X-rays via the $Swift$ satellite, because they were monitoring another SN, SN~2007uy, in the {\emph same} galaxy, when suddenly XRT080109 erupted and lasted $\sim$ 600 sec in X-rays. Furthermore, our program was also monitoring SN 2007uy, and the rest of the host galaxy, in the optical and NIR
from the ground, providing us with stringent limits on the
optical emission just hours before the onset of X-ray transient.

\begin{figure}[!ht]
  \includegraphics[height=.4\textheight]{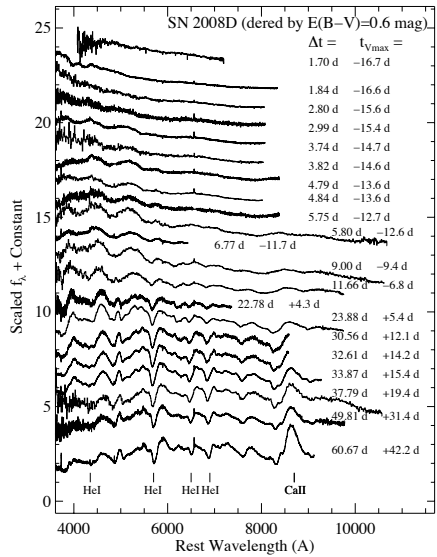} 
  \caption{Spectral evolution of SN 2008D, dereddened by $E(B-V)_{Host}$ = 0.6
mag and labeled with respect to date of shock breakout (indicated by $\Delta$t ), and to date of V-band maximum (indicated by $t_{V max}$). Note the fleeting double-absorption feature around
4000 \AA\ in our early spectrum at $\Delta$t = 1.84 day. The characteristic optical He lines (due to blueshifted He I $\lambda$ 4471, 5876,
6678, 7061) become visible starting t $\sim$12 days or $t_{Vmax} \sim$6 days. From \citet{modjaz09}.}
\label{sn08D-spec-fig}
\end{figure}

\begin{figure}[!ht]
  \includegraphics[height=.45\textheight]{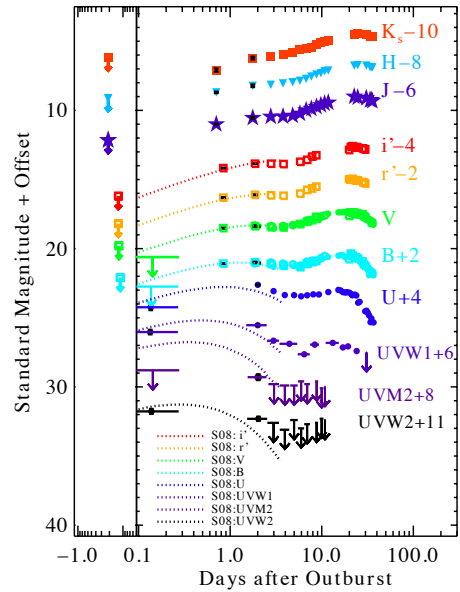} 
  \caption{Observed optical and NIR light curves of SN 2008D after the
onset (the right panel) of XRT 080109, which we adopt as the time of shock breakout. The filled circles show
$UVW1$,$UVM2$, $UVW1$,$U$,$B$, $V$ data from Swift/UVOT, the empty circles are
$BV$ data from KAIT, while
the empty squares are data from the FLWO 1.2 m telescope. $JHKs$ data
(filled stars, triangles, and squares) are from PAIRITEL. Note the very early
optical data points ($\Delta$t =0.84 day after shock breakout) from the FLWO 1.2
m telescope, as well as NIR data (at $\Delta$t =0.71 day) from PAIRITEL, amongst the earliest data of a SN Ib to date. Swift/
UVOT upper limits are indicated by the arrows. We also plot the pre-explosion
upper limits derived from the 1.2m CfA  and the PAIRITEL data (the left panel). The data have not been corrected
for extinction. The  The 2-component light curve is pronounced in the blue, where the 1st peak is due to thermal emission from the expanding and cooling stellar layers and the 2nd peak due to powering by \synNi. We note that our earliest ground-based data points are consistent with the light-curve fits by Soderberg et al. (2008, S08; dotted lines), who use the envelope BB emission model from Waxman et al. (2007). From \citet{modjaz09}.}
\label{sn08D-LC-fig}
\end{figure}

In \citet{modjaz09}, we gathered extensive panchromatic observations (X-ray, UV, Optical, NIR) from 13 different telescopes to determine the nature of SN 2008D, its accompanying Swift X-ray Transient 080109, and its progenitor (see also \citealt{soderberg08}). We first established that SN~2008D is a spectroscopically normal SN Ib (i.e., showing conspicuous He lines, see Fig~\ref{sn08D-spec-fig}), which implies the progenitor star had an intact He layer, but had not retained its outermost H envelope.  For the first time, the very early-time peak (at $\sim 1$~day, see Fig~\ref{sn08D-LC-fig}) could be observed for this kind of SN, from which one can deduce the progenitor radius, since that peak is due to black-body emission from the cooling and expanding stellar envelope. Using our reliable and early-time measurements of the bolometric output of this SN in conjunction with models by \citet{waxman07} and \citet{chevalier08}, as well as published values of kinetic energy and ejecta mass, we derived a progenitor radius of 1.2 $\pm$ 0.7 $R_{\odot}$ (in agreement with \citealt{soderberg08}) and 12 $\pm$ 7 $R_{\odot}$, respectively, the latter being more in line with typical WN stars. We furthermore showed that the observed X-ray emission by which it was discovered \citep{soderberg08} is different from those of X-ray flashes, the weaker cousins of GRBs, which demonstrates that even normal SN Ib, surprisingly, can give rise to high-energy phenomena (but see \citealt{mazzali08}). Lastly, our spectra obtained at three and four months after maximum light show double-peaked oxygen lines that we associate with departures from spherical symmetry, as has been suggested for the inner ejecta of a number of SN Ib cores. Our detailed observations and their analysis, as well as those of others \citep{soderberg08,mazzali08,malesani09,maund09,tanaka09_latetime08d} have inspired a number of theorists to develop and refine sophisticated models of SN shock breakout including relativistic-mediated shocks \citep{katz10},  aspherity \citep{couch11} and the impact of a wind \citep{balberg11} to explain its X-ray shock breakout, as well as refine models of the subsequently cooling envelope \citep{nakar10,rabinak11} aimed at reproducing the observations and predicting their appearance at high-$z$ \citep{tominaga11}.


\section{The Future is Now: The Golden Age of Transient Surveys and Corresponding Host Galaxy Studies}\label{future_sec}

We are embarking on the Golden Age of transients, i.e., diverse explosive phenomena from both massive stars and compact objects, which a number of large-scale surveys are starting to harvest. These innovative surveys include the Palomar Transient Factory (PTF\footnote{http://www.astro.caltech.edu/ptf/}, \citealt{rau09}), Catalina Real-Time Transient Survey (CRTS\footnote{http://crts.caltech.edu/}, \citealt{drake09}), PanSTARRS\footnote{http://ps1sc.org/transients/} \citep{tonry05} and The Chilean Automatic Supernova Search (CHASE\footnote{http://www.das.uchile.cl/proyectoCHASE/}, \citealt{pignata09}, (for a comparison for 2010, see \citealt{gal-yam11}), and in the future, LSST, which is a major US undertaking and number 1 ranked ground-based project during the 2010 Decadal Survey.
Because of different survey modes and detection techniques, these innovative surveys are finding known types of SNe in large numbers and with relatively little bias, as well as rare, yet astrophysically interesting events in statistically large numbers. The main innovations in survey mode  are: very large field-of-view (e.g., 7.8-square-degree for PTF, up to 4000 times larger than traditional surveys), galaxy-untargeted, with different cadences, and less bias towards bright cores of galaxies.

In contrast, traditional SN surveys usually have small fields of view and thus, specifically target
luminous galaxies that contain many stars in order to increase their
odds of finding those that explode as SNe. For example, the prolific Lick Observatory SN Search
(LOSS, \citealt{filippenko01,li11}) monitors a list of galaxies that have a mean (median) value at the
galaxy magnitude of $M_B = -19.9$ ($M_B = -20.1$) mag. But because more luminous galaxies are more metal-rich (\citealt{tremonti04}, see also Fig.~\ref{oxy-modj08-fig}) such targeted SN surveys are probably biased towards finding SN in high-metallicity galaxies. However, the new surveys alleviate this galaxy- and metallicity-bias and PTF has already found over 1090 spectroscopically ID'ed SNe (as of April 2011) in this galaxy-untargeted fashion, probing SNe and transients in all kinds of galactic environments (except in highly obscured ones, of course). The difference in
survey mode appears to be important: using core-collapse SN discovered with PTF,
\citet{arcavi10} showed that different populations of galaxies may be hosting different types of stripped SNe.
The least stripped SN (SN IIb) were found in the low luminosity galaxies (dwarfs), which Arcavi et al. took to indicate a metallicity effect, where the massive progenitor at low-metallicity did not have sufficiently strong winds to remove its He layer in order to explode as a SN Ic. While these findings are in line with \citet{modjaz11}, in order to verify them with direct metallicity measurements, as well as for resolving some of the issues discussed in Section~\ref{ccsnz_sec}, it is necessary to conduct a thorough and extensive host galaxy study with a large single-survey, untargeted, spectroscopically classified, and homogeneous collection of stripped SNs, something we are currently undertaking with PTF.

For core-collapse SNe and transients, the next big frontier is to hunt for the lowest metallicity host galaxies, in order to fully confront theoretical predictions of the impact of metallicity on stellar death \citep{heger03,oconnor11} with observations, via the untargetted surveys or specifically low-luminosity galaxy-targeted surveys.

\acknowledgements
I thank the Astronomische Gesellschaft (German Astronomical Society) and its selection committee for awarding me the 2010 Ludwig-Biermann Award and for giving me the opportunity to participate in the 2010 annual meeting in Bonn. Furthermore, I thank my many collaborators over the years and at various institutions for their fruitful collaboration. M.M. acknowledges support by the Hubble Fellowship grant
HST-HF-51277.01-A, awarded by STScI, which is operated by AURA under
NASA contract NAS5-26555. This research has made use of NASAÕs Astrophysics Data System.

\bibliographystyle{apj}
\bibliography{/Users/maryammodjaz/Dropbox/refs}


\clearpage



\newpage



\end{document}